\documentclass{article}
\usepackage{PRIMEarxiv}

\usepackage[utf8]{inputenc} 
\usepackage[T1]{fontenc}    
\usepackage{hyperref}       
\usepackage{url}            
\usepackage{booktabs}       
\usepackage{amsfonts}       
\usepackage{nicefrac}       
\usepackage{microtype}      
\usepackage{lipsum}
\usepackage{fancyhdr}       
\usepackage{graphicx}       
\graphicspath{{media/}}     

\title{{Pb Substitution Effects on Lattice and Electronic System of the BiS$_{2}$-based Superconductors La(OF)BiS$_{2}$}

\author{
\\Miku Sasaki , Kotaro Inada , Fumito Mori , Takaaki Hirase , Haruki, Yamada , \\
Shota Shimoyama , Yasushi Nakamura , Tetta Nakamura , \\
Yoshiyuki Shibayama , and Naoki Momono
\\[1ex]
Muroran Institute of Technology, Muroran, Hokkaido 050-8585, Japan \\
} 

}

\begin{document}
\maketitle
\begin{abstract}
We examined the effect of Pb substitution in the layered superconductor $\mathrm{La}$$\mathrm{O}_{0.5}$$\mathrm{F}_{0.5}$$\mathrm{Bi}_{1-x}$$
\mathrm{Pb}_{x}$$\mathrm{S}_{2}$ (${x}=0{\sim}0.15$) through the measurements of the resistivity, thermal expansion, specific heat, and Seebeck coefficient. These transport and thermal properties show anomalies at certain temperatures (T*) for ${x}{\geq}0.08$. The large thermal expansion anomalies, specific heat anomalies, and the existence of hystereses in the above measurements indicate a first-order structural phase transition at T*.  Additionally, the Seebeck coefficient indicates that the anomalies at T* are related not only to the lattice system, but also to the electronic system. Superconductivity is not observed above 2 K at ${x}$=0.08, which is around the phase boundary where T* vanishes. The suppression of superconductivity around the structural phase boundary suggests a close relationship between the lattice and superconductivity.
\end{abstract}


\section{Introduction}
Since the BiCh$_{2}$-based (Ch: chalcogenide) superconductor Bi$_{4}$O$_{4}$S$_{3}$ was discovered in 2012\cite{1}, various BiCh$_{2}$-based superconductors such as La(O, F)BiCh$_{2}$ (Ch=S, Se) have been studied intensively  \cite{2,3,4,5,6,7,8,9,10,11}. 
A major feature of BiCh$_{2}$ superconductors is absence of the isotope effect on the superconducting transition temperature $T_{c}$ ; the isotope exponent ${\alpha}$ defined by $T_{c}$ ${\sim}$ M$^{\alpha}$ (M : the isotope mass) was close to zero in 
LaO$_{0.6}$F$_{0.4}$BiSSe and Bi$_{4}$O$_{4}$S$_{3}$, which is expected to be ${\alpha}$ ${\sim}$0.5 in conventional BCS superconductors \cite{12,13}. 
Another feature is that in LaO$_{1-{x}}$F$_{x}$BiSSe  and CeOBiS$_{1.7}$Se$_{0.3}$ crystals with tetragonal structure, $c$-axis magnetoresistance in the superconducting state shows two-fold symmetry in in-plane anisotropy \cite{14,15,16}. These results suggest that BiCh$_{2}$-based superconductors have an unconventional superconducting mechanism. 

BiCh$_{2}$-based superconductors La(O, F)BiCh$_{2}$ have the layered structures comprising of block layers (LaO) and conduction layers (BiCh$_{2}$). By partially substituting F for O in the block layers, we can introduce electronic carriers into the conduction layers, leading to the emergence of superconductivity \cite{10}. Interestingly, $T_{c}$ increases significantly by applying pressure on the crystals. LaO$_{0.5}$F$_{0.5}$BiS$_{2}$ has $T_{c}$ ${\sim}$ 2 K under ambient pressure \cite{17}. 
 By applying a pressure of 1 GPa, $T_{c}$ increases up to ${\sim}$10 K ,\cite{18, 19} accompanied by a change from the tetragonal structure (${P}$4/${nmm}$) to a monoclinic one (${P}$2$_{1}$/${m}$) . 
These results indicate that the crystal structure may be responsible for the increase in $T_{c}$ \cite{20}.  Additionally chemical pressure introduced instead of physical pressure, is also effective in increasing $T_{c}$ \cite{21}. Mizuguchi et al. reported that an in-plane chemical pressure suppresses in-plane disorders and enhances the lattice anharmonicity, which is positively linked to $T_{c}$ and the pairing mechanism\cite{22}.

Recently, studying the effect of chemical pressure on Pb-doped BiCh$_{2}$ crystals has garnered significant attention, which cannot be understood solely based on the in-plane chemical pressure effects.  In NdO$_{0.7}$F$_{0.3}$Bi$_{1-{x}}$Pb${x}$S$_{2}$, the length of the ${c}$-axis decreases monotonically with the Pb concentration (${x}$) until ${x}$=0.06, and $T_{c}$ increases up to 5.6 K for ${x}$=0.06 \cite{23}. Moreover, in LaO$_{0.5}$F$_{0.5}$Bi$_{1-{x}}$Pb$_{x}$S$_{2}$, $T_{c}$ exhibits a suppression between ${x}$=0.02 and 0.05 and an enhancement above ${x}$=0.06. Interestingly, LaO$_{0.5}$F$_{0.5}$Bi$_{1-{x}}$Pb$_{x}$S$_{2}$ shows an anomaly in the temperature dependence of the resistivity at approximately 150 K above ${x}$=0.08 \cite{24}. This anomaly was not observed in NdO$_{0.7}$F$_{0.3}$Bi$_{1-{x}}$Pb${x}$S$_{2}$. The origin of this anomaly and its relation to the increase in $T_{c}$ are still not clear.

In this study, we examined the effect of Pb-substitution on the resistivity, thermal expansion, specific heat, and Seebeck coefficient for LaO$_{0.5}$F$_{0.5}$Bi$_{1-{x}}$Pb$_{x}$S$_{2}$ (${x}$ =0${\sim}$0.15) single crystals. We observed anomalies and hystereses at a certain temperature, T*, above ${x}$=0.08 in all transport and thermal measurements, and conclude that these anomalies at T* are due to the first-order structural phase transition. The superconductivity is suppressed at ${x}$=0.08 around the boundary where T* vanishes. 

\section{Experiment}
Single crystalline samples of LaO$_{0.5}$F$_{0.5}$Bi$_{1-{x}}$Pb$_{x}$S$_{2}$ (${x}$ =0${\sim}$0.15) were synthesized with a flux method with KCl and CsCl in a quartz tube. Powders of La$_{2}$S$_{3}$, Bi$_{2}$O$_{3}$, Bi$_{2}$S$_{3}$, PbF$_{2}$, BiF$_{3}$, and Bi were used as starting materials. The molar ratio of the KCl and CsCl flux was KCl : CsCl$=3:5$. The powders (2 g) and flux (15 g) were mixed and vacuum-sealed in a quartz tube. The tube was heated up to 900 ${}^\circ$C for 12 hours, maintained at 900 ${}^\circ$C for 24 hours, and then cooled down to 600 ${}^\circ$C at a rate of 0.5 ${}^\circ$C/h.

 The resistivity was measured using PPMS (Physical Property Measurement System, Quantum Design) with the conventional four-terminal method from 2 K to 300 K. The thermal expansion was measured using a self-made device in the laboratory with the two-gauge method from 80 K to 300 K. The specific heat was measured using chopped-light AC calorimetry (ADVANCE RIKO) from 85 K to 300 K. We also measured specific heat using PPMS from 0.38 K to 300 K. The Seebeck coefficient was measured from 4 K to 300 K using a device constructed in the laboratory. 

Single crystal X-ray structural analysis was conducted using a Rigaku VariMax with Saturn at 293 K. The radiation source was MoK$_{\alpha}$ (${\lambda}$ = 0.71073 \mbox{\AA}). Since O and F are not distinguishable by X-ray diffraction, F was not included in the molecular model considered. Structural analysis was conducted with the Bi and Pb contents fixed. From single crystal X-ray structural analysis, we obtained the space group of ${P}$4/${nmm}$ ({\#}129) for all Pb concentrations. The parameters R1, wR2, and goodness of fit indicator are 0.0396 to 0.0637, 0.0896 to 0.1494, and 1.146 to 1.287, respectively. We examined the Pb concentration using an electron probe microanalyzer (EPMA). The obtained Pb concentration increased linearly with the nominal Pb concentration; the net Pb concentration was 17 {\%} higher than the nominal Pb concentration. We use the nominal Pb concentration to indicate the crystals, and the net Pb concentration was obtained as 1.17${x}$, where ${x}$ is the nominal concentration.

\section{Results and Discussion}

\subsection{Crystal structure and F concentration}
 Figures 1 (a) and (b) show the dependences of lattice parameter ${a}$ and ${c}$-axis on the Pb concentration at 293 K. The ${a}$ and ${c}$-axis increase with Pb concentration until ${x}$=0.14 and decrease above ${x}$=0.15. The increase in ${a}$ and ${c}$ can be ascribed to the difference between the ionic radii of Pb$^{2+}$ and Bi$^{2.5{\sim}3+}$; ${R}$ (Pb$^{2+}$) =1.19 ${\mbox{\AA}}$ \cite{25}, ${R}$ (Bi$^{3+}$) =1.03 ${\mbox{\AA}}$\cite{25}, and ${R}$  (Bi$^{2.5+}$) =1.0419 ${\mbox{\AA}}$ \cite{21}. 
The lattice parameter ${a}$ and ${c}$ axis are ${a}$ =4.0544(10) ${\mbox{\AA}}$, ${c}$ =13.4729(3) ${\mbox{\AA}}$ for LaO$_{0.5}$F$_{0.5}$BiS$_{2}$, respectively.  The detailed dependence of the crystal structure on the Pb concentration will be discussed in another paper, in which we will report that the crystal has monoclinic structure (${P}$2$_{1}$/${m}$) at room temperature above ${x}$=0.18.

In this study, O: F= 0.5: 0.5 in nominal composition; the F content cannot be determined precisely using X-ray analysis, because there is negligible difference between the scattering factors of O and F. It was reported that the length of the ${c}$-axis monotonously decreases with increasing ${y}$ in LaO$_{1-{y}}$F$_{y}$BiS$_{2}$\cite{17,26}. We estimated the net F concentration of the crystals in this study by comparing our data with the ${c}$-axis length reported in Ref. 17, resulting in an F concentration of ${\sim}$0.35. 
As shown in Ref. 17, the ${c}$-axis of the pure crystal increased by approximately 0.1 ${\AA}$ per 10 {\%} F concentration. In the present Pb-doped crystal, the ${c}$-axis increases slightly with ${x}$ from ${x}$=0 to 0.14 by approximately 0.02 ${\mbox{\AA}}$. We exclude the relatively large change in the ${a}$ and ${c}$-axes for ${x}$=0.15, because these changes occur near the phase boundary discussed later and its origin seems to be different from the change in ${a}$ and ${c}$-axis from ${x}$=0 to 0.14. Assuming that the change in the ${c}$-axis is cased due to the F concentration only, the change in the F concentration is only 2 {\%} between ${x}$=0${\sim}$0.14. Then, we conclude the F concentrations of the present Pb-doped crystals is approximately 0.33${\sim}$0.37. The length of the ${a}$ and ${c}$-axes in the present Pb-doped crystals is different from those reported in Otsuki et al \cite{24}, which may be ascribed to the difference in the F concentration.

\subsection{Resistivity}
Figures 2 (a)-(c) show the ${T}$-dependence of the resistivity (${\rho}$) in the ${T}$ range from 2 K to 300 K for LaO$_{0.5}$F$_{0.5}$Bi$_{1-{x}}$Pb$_{x}$S$_{2}$  (${x}$= 0${\sim}$0.12). Fig.2 (a) shows ${\rho}$ in the cooling process for ${x}$${<}$0.08; ${\rho}$ exhibits a semiconducting behavior with the Pb concentration until ${x}$=0.06. These semiconducting behaviors are similar to that reported by Otsuki et al \cite{24}. For ${x}$=0${\sim}$0.07, ${\rho}$ starts to decrease around 2${\sim}$3 K due to superconductivity. As shown in Fig. 2 (b), ${\rho}$ increases with decreasing ${T}$, and drops sharply at a certain temperature T*$_{ER}$. The sharp change in ${\rho}$ often appears several times below T*$_{ER}$. The ${\rho}$ below T*$_{ER}$ shows a metallic behavior except for the low temperatures of ${T}$${<}$30 K. Such a change in ${\rho}$ around T*$_{ER}$ has been reported by Otsuki et al \cite{24}. For ${x}$=0.09 and 0.10, ${\rho}$ started to drop due to superconductivity at approximately 3 K. In this study, we evaluated ${T}_{c}^{ER onset}$ from the temperature at which ${\rho}$ starts to decrease sharply, because zero resistance was not observed for all Pb concentrations.
  
For ${x}$=0.08, ${\rho}$ shows a sharp peak at ${\sim}$10 K, although its origin is still not clear. Figure 2 (c) shows ${\rho}$ in the heating process; ${\rho}$ shows a similar behavior to that in the cooling process, although T*$_{ER}$ in the heating process is approximately 15-32 K higher than that in the cooling process. This behavior of ${\rho}$ indicates that the change in ${\rho}$ at T*$_{ER}$ is accompanied by hysteresis, suggesting a first-order phase transition.

\subsection{Thermal expansion}
Figures 3 (a)-(c) show the ${T}$-dependence of the linear expansion coefficient  ${\Delta}{L}$(${T}$)/${L}$ in the ${T}$ range from 85 K to 300 K for LaO$_{0.5}$F$_{0.5}$Bi$_{1-{x}}$Pb$_{x}$S$_{2}$ (${x}$=0.08${\sim}$0.14). Here, ${\Delta}{L}$(${T}$)/${L}$ was obtained by integrating the thermal expansion coefficient, ${\alpha}$, with respect to ${T}$. ${\Delta}{L}$(${T}$)/${L}$ at ${x}$=0.10 and 0.14 decreases with decreasing ${T}$ at a nearly constant rate, and drops sharply around T*$_{ER}$.  After the sharp drops,  ${\Delta}{L}$(${T}$)/${L}$ decreases again with ${T}$. For ${x}$=0.10, the overall ${T}$-dependence of  ${\Delta}{L}$(${T}$)/${L}$ is similar to those of other crystals except for the anomaly, where  ${\Delta}{L}$(${T}$)/${L}$ for ${x}$=0.10 shows a small peak. The anomaly of  ${\Delta}{L}$(${T}$)/${L}$ shows hysteresis (Figure 3(b)). Insets of Figures 3 show the thermal expansion coefficient, ${\alpha}$, which shows a large peak corresponding to the sharp change of  ${\Delta}{L}$(${T}$)/${L}$ originating from the change in the lattice structure.  We define T*$_{TE}$ as the peak temperature of ${\alpha}$. The anomalies of  ${\Delta}{L}$(${T}$)/${L}$ and ${\alpha}$ accompanied by hysteresis indicate that the change in ${\rho}$ at T*$_{ER}$ arises from the first-order phase transition of the lattice system.

\subsection{Specific heat}

Figure 4 (a) shows the specific heat, ${C}$, of LaO$_{0.5}$F$_{0.5}$Bi$_{1-{x}}$Pb$_{x}$S$_{2}$ (${x}$=0.09) in the ${T}$ range from 10 K to 300 K. Figure 4 (b) shows the ${C}$/${T}$ vs. ${T}^{2}$ plot for ${x}$=0.09 from 0.38 K to 5 K. In Fig. 4 (b), the dashed line represents the linear fit of the data with ${C}$/${T}$=${\gamma}_{N}$ + ${\beta}$${T}$$^{2}$, where ${\gamma}_{N}$ is the electronic specific heat coefficient and ${\beta}$ is the coefficient of the phonon cubic term. We obtained ${\gamma}_{N}$=1.25 mJ/mol·K$^{2}$ and ${\beta}$=0.62 mJ/mol·K$^{4}$ from the fitting. The transition between the superconducting state and normal state at approximately ${T}$ ${\sim}$3.5 K is not distinct. Inset of Fig. 4 (b) shows the electronic specific heat, ${C}_{e}$, divided by ${T}$ (${C}_{e}$/${T}$), which is obtained by subtracting the phonon contribution ${\beta}$${T}$$^{2}$ from ${C}$/${T}$ (${C}_{e}$/${T}$ =${C}$/${T}$-${\beta}$${T}$$^{2}$) under the assumption that the phonon terms in the normal state do not change in the superconducting state. There is a hump at approximately 3 K in the inset of Fig. 4 (b); ${C}_{e}$/${T}$ starts to increase below 3.4 K(= $T_{c}^{SH onset}$ ) and decreases toward zero below ${\sim}$ 2 K. $T_{c}^{SH onset}$  is close to $T_{c}^{ER onset}$, which is defined in terms of the resistivity. Despite the broad superconducting transition, the low-temperature specific heat, ${C}_{e}$/${T}$, approaches zero. This behavior suggests zero residual ${\gamma}$; and specifically, the bulk nature of superconductivity for at least ${x}$=0.09. 

Figures 5 (a) and (b) show the ${C}$ in the ${T}$ range from 80 K to 300 K for LaO$_{0.5}$F$_{0.5}$Bi$_{1-{x}}$Pb$_{x}$S$_{2}$ (${x}$=0.09 and 0.10). In Figs. 5 (a) and (b), ${C}$ is not the absolute value, but relative value in arbitrary units, because the values were measured by AC calorimetry. The ${C}$ values for ${x}$=0.09 and 0.10 slowly decrease with decreasing ${T}$ and drop sharply at T*$_{SH}$. The anomalies in ${C}$ for ${x}$=0.09 and 0.10 show  hystereses as well as ${\rho}$ and ${\Delta}{L}$(${T}$)/${L}$, indicating a first-order phase transition. Here, we focus on the magnitude of the anomaly in ${C}$, ${\Delta}{C}$. ${\Delta}{C}$/${T}$ for ${x}$=0.09 and 0.10 is 25 and 109 mJ/mol·K$^{2}$, which are estimated from the comparison of the ${C}$ data measured by PPMS and AC calorimetry. The ${\Delta}{C}$/${T}$ values are significantly higher than the ${\gamma}_{N}$ value of 1.25 mJ/mol·K$^{2}$. This implies that the contribution of the electronic system to ${\Delta}{C}$ is small, and the transition at T*$_{SH}$ is mostly caused by the lattice system. 

\subsection{Seebeck coefficient}
Figures 6 (a)-(c) show the ${T}$-dependence of the Seebeck coefficient ${S}$ in the ${T}$ range from 4 K to 300 K for  LaO$_{0.5}$F$_{0.5}$Bi$_{1-{x}}$Pb$_{x}$S$_{2}$  (${x}$ = 0${\sim}$0.10). For ${x}$=0, ${S}$ shows a ${T}$-linear dependence above ${\sim}$200 K and significantly increases below ${\sim}$ 200 K. ${S}$ has a broad peak at ${\sim}$30 K and decreases toward zero at lower temperatures. For ${x}$=0.02 and 0.04, ${S}$ increases significantly below ${\sim}$210 K and ${\sim}$ 240 K and shows a peak around ${\sim}$30 K. These behaviors of ${S}$ observed for ${x}$=0.02 and 0.04 are similar to those of ${x}$=0. For ${x}$=0.07 and 0.08; ${S}$ increases below ${\sim}$ 240 K and ${\sim}$ 230 K, but the enhancement becomes smaller than that for ${x}$=0.04. ${S}$ for ${x}$=0.08 changes sign below ${\sim}$10 K, which may be related to the sharp peak of ${\rho}$.

The temperature (${T}_{p}$) where ${S}$ starts to deviate from ${T}$-linear dependence at high temperatures is close to the Debye temperature determined by phonon specific heat coefficient ${\beta}$ \cite{27}, as shown in the inset of Fig. 6 (b).  The significant enhancement of ${S}$ below ${T}_{p}$ can be ascribed to a positive phonon-drag effect. The large positive phonon-drag effects for ${x}$${\leq}$0.08 indicate strong phonon-electron interactions and strong Umkrapp scattering enhanced by Pb substitution.

For ${x}$=0.09 and 0.10, ${S}$ is approximately constant at high temperatures (${T}$${>}$170 K for ${x}$=0.09 and 250 K for ${x}$=0.10)  and shows a step-like anomaly at T*$_{Seebeck}$, which is approximately 160 K and 200 K for ${x}$=0.09 and 0.10, respectively. ${S}$ is approximately proportional to ${T}$ below ${\sim}$100 K. Interestingly, the positive phonon-drag enhancement in ${S}$ was not observed for ${x}$=0.09 and 0.10. Instead, ${S}$ for ${x}$=0.09 and 0.10 shows a weak negative phonon-drag effect at low temperatures of 10-30 K. This result means that the nature of the phonon-electron interactions changes below and above ${x}$=0.08.

The Seebeck coefficient is generally not much affected by structural phase transitions. It seems that the anomalies at T* are related not only to the lattice system, but also to the electronic system. From first-principle band structure calculations, Kuroki et al. reported that the phase transition from the tetragonal to monoclinic structure for LaO$_{1-{y}}$F$_{y}$BiS$_{2}$ under high pressure causes a band splitting, which leads to a decrease in the electronic energy in monoclinic structure \cite{28}. The Seebeck coefficient anomalies could be ascribed to a band splitting in the electronic system due to the structural phase transition from the tetragonal to monoclinic structures.

\subsection{Pb dependences of ${T}_{c}$ and T*}
Figure 7 shows the Pb dependence of T*, obtained by the various measurements in LaO$_{0.5}$F$_{0.5}$Bi$_{1-{x}}$Pb$_{x}$S$_{2}$ (${x}$=0${\sim}$0.15). ${T}_{c}^{onset}$ obtained from the resistivity and specific heat are also plotted together with T* in Fig. 7.  ${T}_{c}^{ER onset}$ approximately agrees with ${T}_{c}^{SH onset}$ at least for ${x}$=0 and 0.09 \cite{27}. ${T}_{c}^{ER onset}$ is nearly independent of the Pb concentration except for that at ${x}$=0.08; there is no sign of superconductivity at ${x}$=0.08 at least above ${\sim}$2 K. It should be noted that the suppression of ${T}_{c}$ at ${x}$=0.08 is in contrast to the results reported by Otsuki et al; this may be due to the slight difference in the F concentration, as described previously in Section 2.  Moreover, the measured  T*$_{ER}$ in this study is consistent with data reported by Otsuki et al. T* seems to be independent of the F concentration, that is, the electronic doping level. 

As shown in Fig. 7, T* appears around ${x}$=0.08 and increases linearly with the Pb concentration. The dependences of T* on the Pb concentration measured by various probes are similar to each other. 
The specific heat anomaly and existence of hystereses in the measurements indicate the first-order phase transition at T*, as described before. Here, we define the Low-${T}$ phase as the temperature region below T* and High-${T}$ phase as the temperature region above T*. At ${x}$=0.08, which is just the boundary between High-${T}$ and Low-${T}$ phases, superconductivity is not observed above 2 K. In other words, superconductivity is suppressed near the phase boundary. The large anomaly in thermal expansion suggests some type of structural phase transition from the tetragonal to another structure at T*. 
The suppression of superconductivity at the structural phase boundary suggests the close relationship between the lattice structure and superconductivity.

\section{Conclusion}
In conclusion, we have measured the resistivity, thermal expansion, specific heat, and Seebeck coefficient in a single crystal of LaO$_{0.5}$F$_{0.5}$Bi$_{1-{x}}$Pb$_{x}$S$_{2}$ (${x}$=0${\sim}$0.15). The dependences of T* on the Pb concentration measured by various probes are similar to each other. The large anomaly of thermal expansion, shape of the specific heat anomaly, and the existence of hystereses indicate a first-order structural phase transition at T* from the tetragonal symmetry (${P}$4/${nmm}$). The suppression of ${T}_{c}$ at ${x}$=0.08 near the phase boundary suggests the close relationship between the lattice structure and superconductivity.

\newpage
\begin{figure}
  \centering
  \includegraphics[keepaspectratio,scale=0.4]{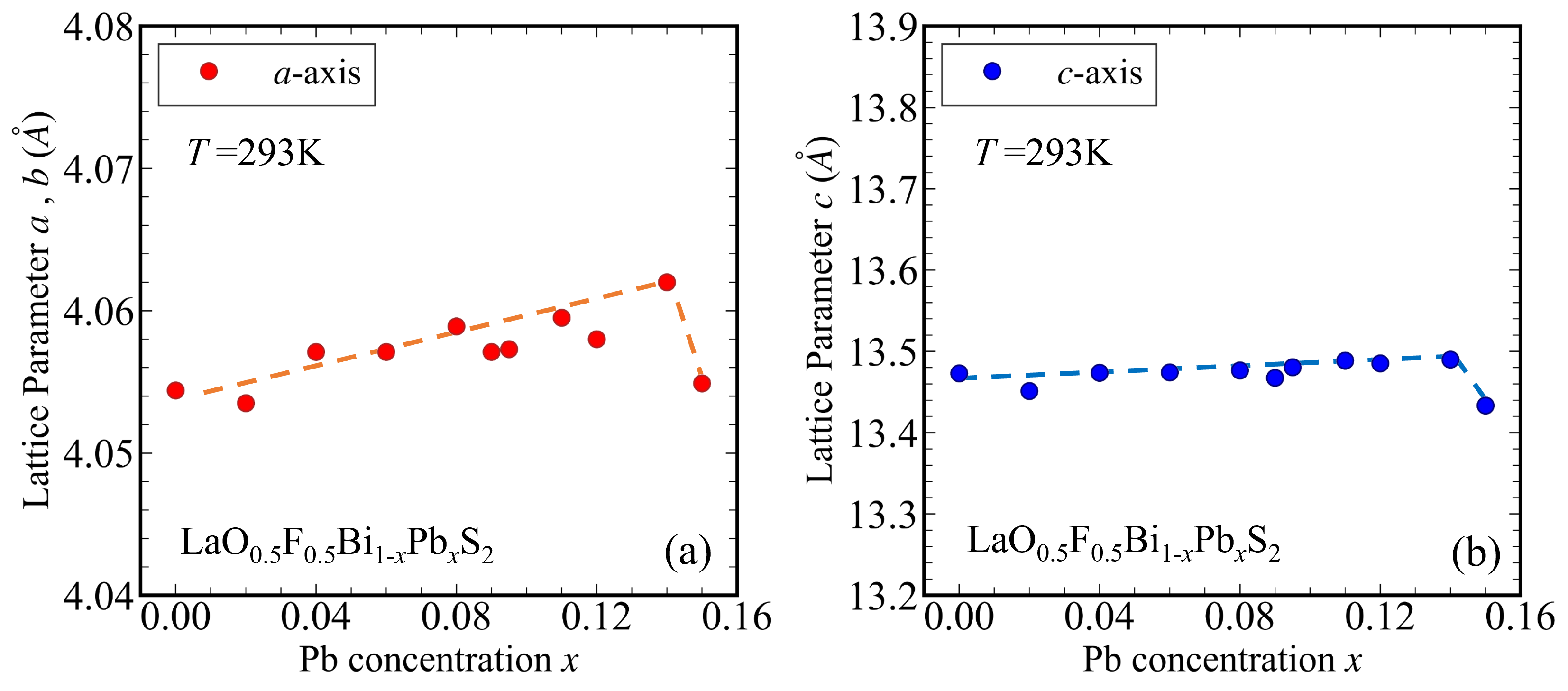}
  \caption{(a) Pb dependence of the lattice parameter ${a}$ (red circle) and (b) lattice parameter ${c}$ (blue circle) for LaO$_{0.5}$F$_{0.5}$Bi$_{1-{x}}$Pb$_{x}$S$_{2}$(${x}=0 {\sim} 0.15$) at 293 K. The red and blue dashed line are guide to eye. }
  \label{f1}
\end{figure}

\newpage
\begin{figure}
  \centering
  \includegraphics[keepaspectratio,scale=0.4]{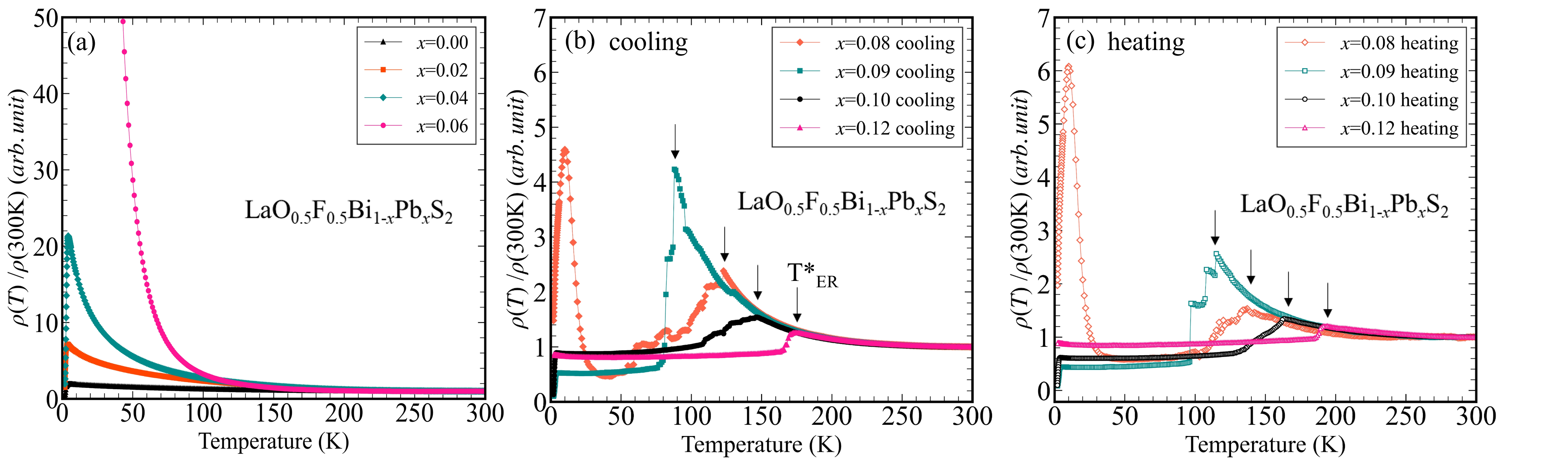}
  \caption{(a) Temperature dependence of the electrical resistivity ${\rho}$(${T}$) / ${\rho}$(300 K) for LaO$_{0.5}$F$_{0.5}$Bi$_{1-{x}}$Pb$_{x}$S$_{2}$(${x}$=0${\sim}$0.06) in the cooling process. The data for ${x}$=0, 0.02, 0.04, 0.06 are shown by black (triangle), orange (scares), cyan (rhombus) and pink (circle). (b) and (c) Temperature dependence of the electrical resistivity ${\rho}$(${T}$) / ${\rho}$(300 K) for LaO$_{0.5}$F$_{0.5}$Bi$_{1-{x}}$Pb$_{x}$S$_{2}$(${x}$=0.08, 0.09, 0.10, 0.12) from 2 K to 300 K. The data for ${x}$=0.08, 0.09, 0.10 and 0.12 are shown by orange (rhombus), cyan (scares), black (circles) and pink (triangles) in (b) and (c), respectively. The cooling (solid symbols) and heating process (open symbols) are shown in (b) and (c), respectively.}
  \label{f2}
\end{figure}

\newpage
\begin{figure}
  \centering
  \includegraphics[keepaspectratio,scale=0.55]{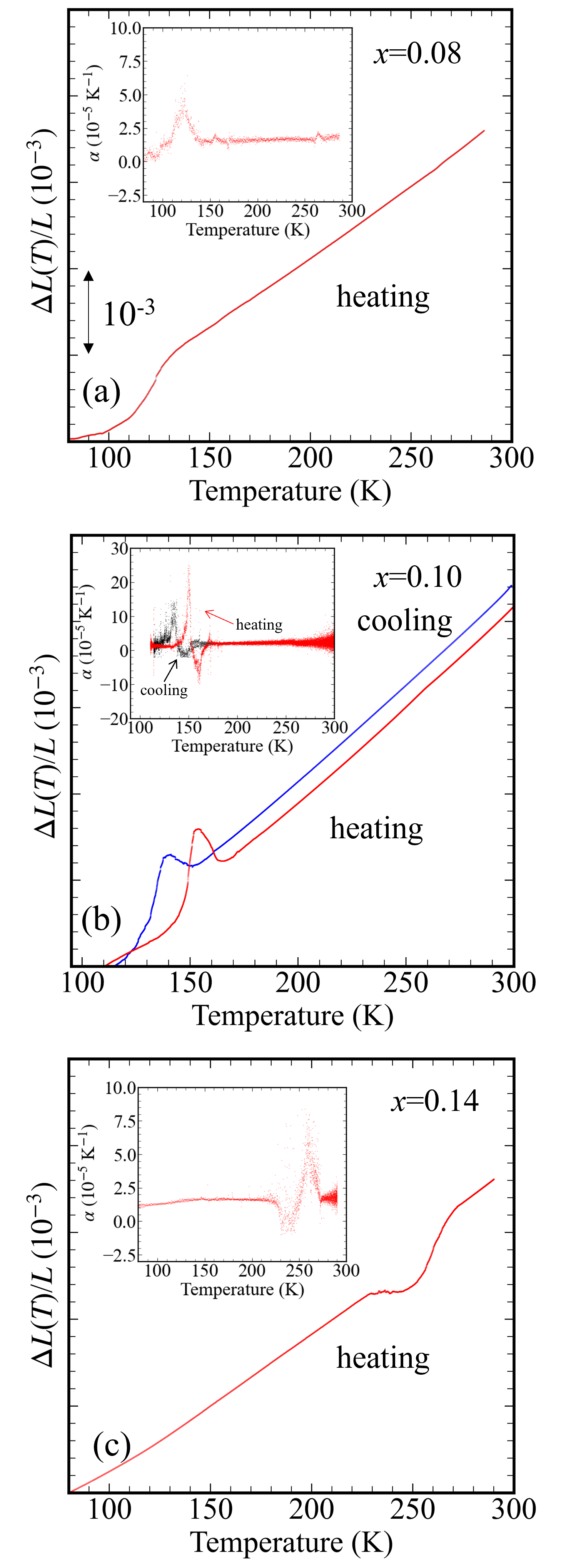}
\caption{(a)-(c) Temperature dependence of the linear expansion ${\Delta}{L}$(${T}$)/${L}$, obtained by integrating the thermal expansion coefficient (${\alpha}$) with respect to ${T}$. Inset figure shows the thermal expansion coefficient for LaO$_{0.5}$F$_{0.5}$Bi$_{1-{x}}$Pb$_{x}$S$_{2}$ (${x}$=0.08, 0.10, 0.14) from 85 K to 300 K. Red and blue (Inset figure is black) lines represent the heating and cooling processes respectively. }
  \label{f3}
\end{figure}

\newpage
\begin{figure}
  \centering
  \includegraphics[keepaspectratio,scale=0.55]{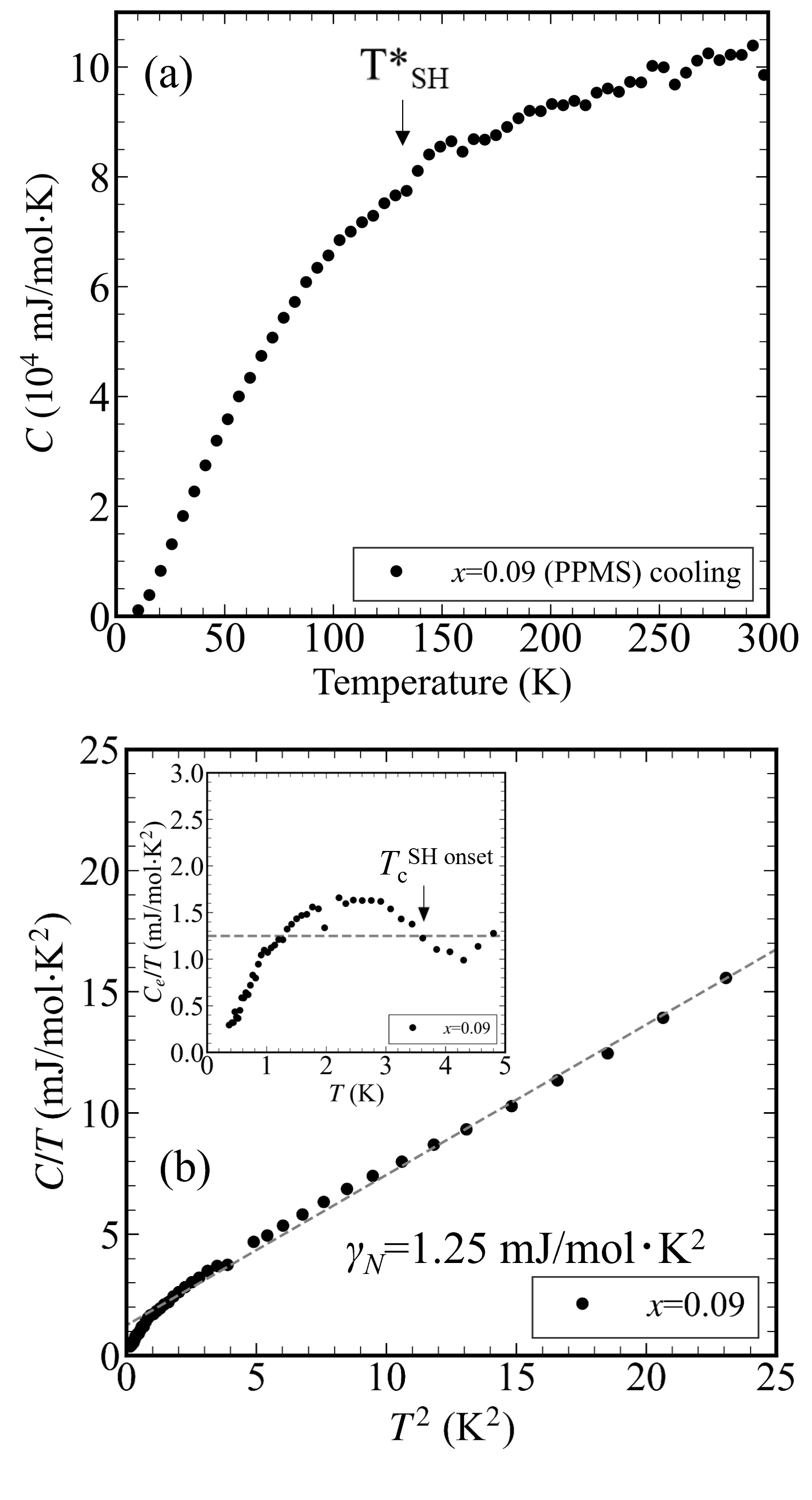}
  \caption{(a) Temperature dependence of the specific heat (${C}$) for LaO$_{0.5}$F$_{0.5}$Bi$_{1-{x}}$Pb$_{x}$S$_{2}$ (${x}$=0.09) from 10 K to 300 K measured using PPMS. (b) ${C}$/${T}$ vs. ${T}$$^{2}$  plot of LaO$_{0.5}$F$_{0.5}$Bi$_{1-{x}}$Pb$_{x}$S$_{2}$ (${x}$=0.09) from 0.38 K to 5 K. The dashed line represents the linear fit of the data with ${C}$/${T}$=${\gamma}_{N}$ + ${\beta}$${T}$$^{2}$. Inset figure shows ${C}_{e}$/${T}$ vs ${T}$ plot around ${T}_{c}$. The dashed line is the ${\gamma}_{N}$ value.}
  \label{f4}
\end{figure}

\newpage
\begin{figure}
  \centering
  \includegraphics[keepaspectratio,scale=0.55]{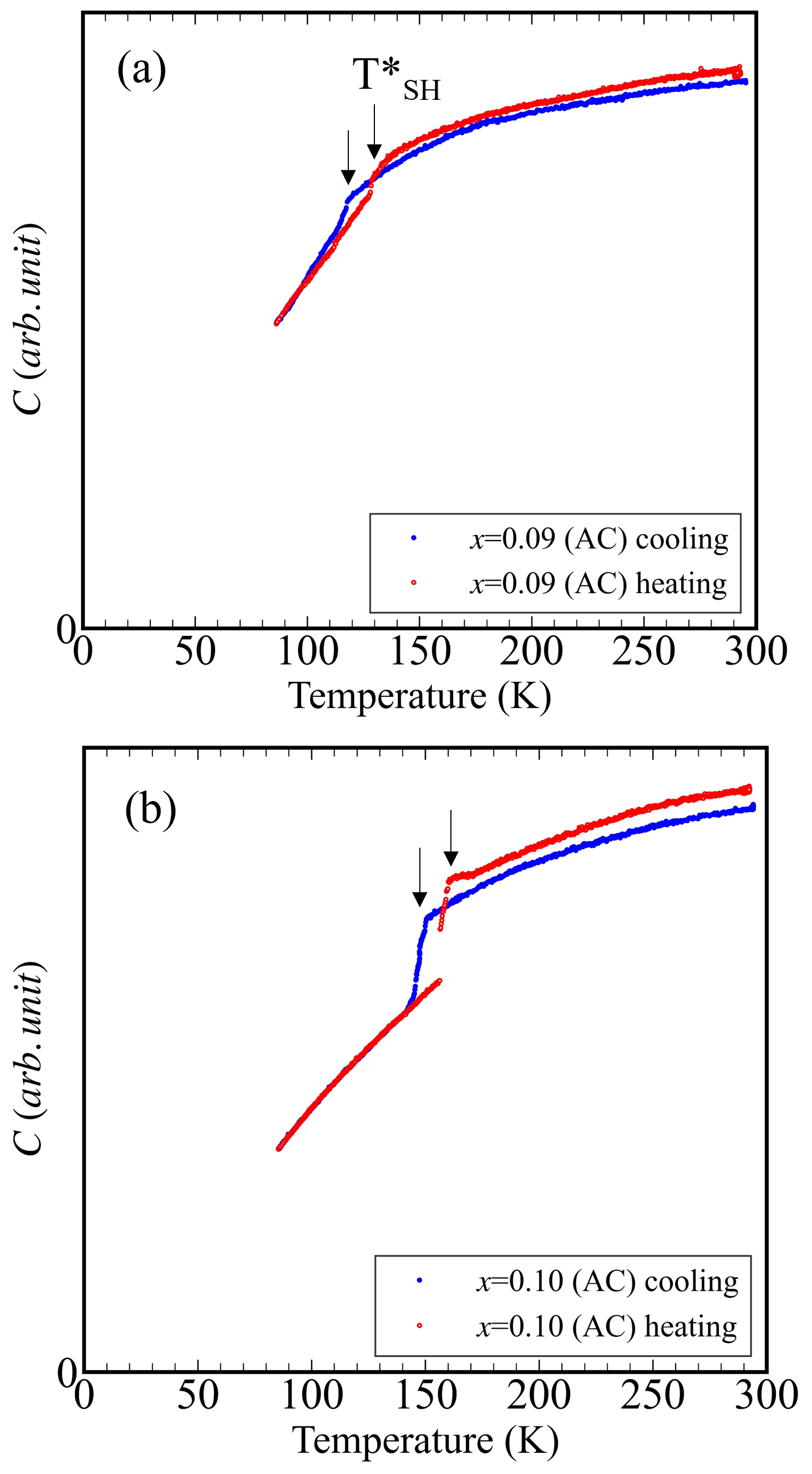}
  \caption{Temperature dependence of the specific heat for LaO$_{0.5}$F$_{0.5}$Bi$_{1-{x}}$Pb$_{x}$S$_{2}$  (${x}$=0.09 and 0.10) from 80 K to 300 K measured using chopped-light AC calorimetry. The red open circles indicates the data in heating process and the blue closed circles represents the data on cooling process. }
  \label{f5}
\end{figure}

\newpage
\begin{figure}
  \centering
  \includegraphics[keepaspectratio,scale=0.6]{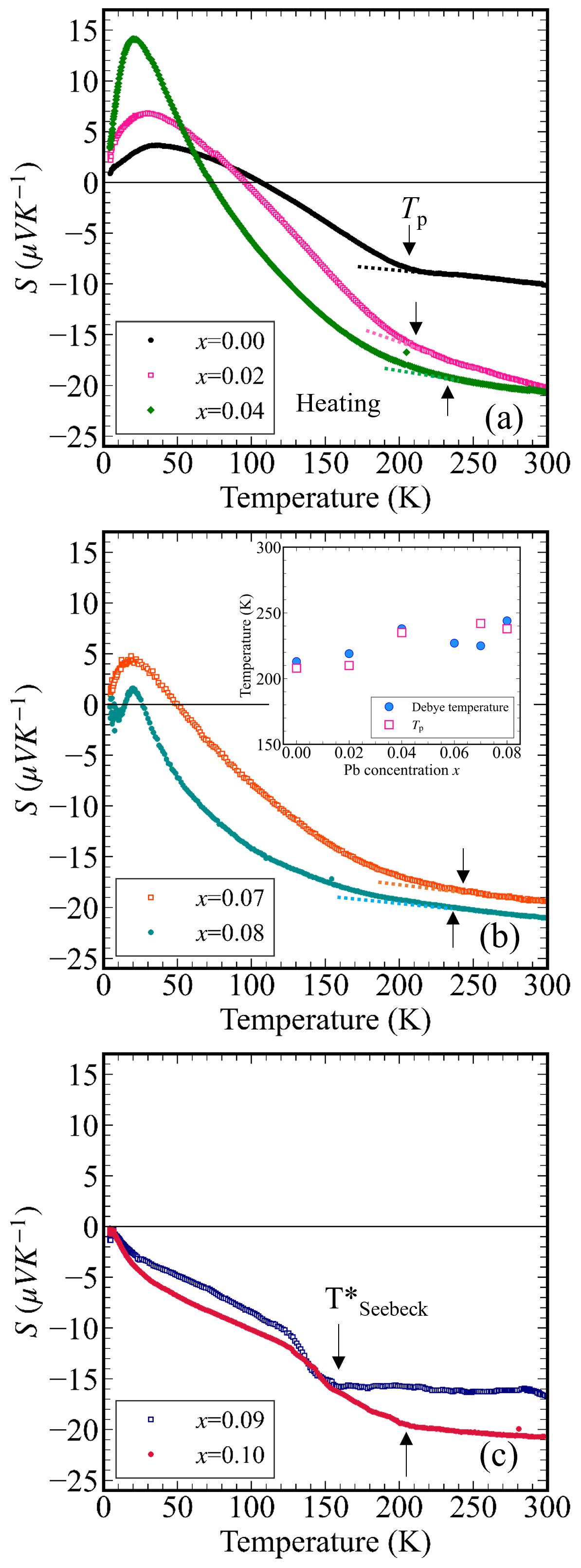}
  \caption{(a)-(c) Temperature dependence of the Seebeck coefficient (${S}$) for LaO$_{0.5}$F$_{0.5}$Bi$_{1-{x}}$Pb$_{x}$S$_{2}$ (${x}$=0${\sim}$0.10) from 4 K to 300 K in the heating process. The data for ${x}$=0.00, 0.02, 0.04, 0.07, 0.08, 0.09 and 0.10 are shown by black (circles), pink (open screals), green (rhombus), orange (open screals), cyan (circles), blue (open screals) and red (circles) in (a), (b) and (c), respectively. Inset figure shows the Pb dependence of the Debye temperature (blue closed circles) and temperature ${T}_{p}$ (pink open squares) from which the positive phonon-drag effects become marked. }
  \label{f6}
\end{figure}

\newpage
\begin{figure}
  \centering
  \includegraphics[keepaspectratio,scale=0.6]{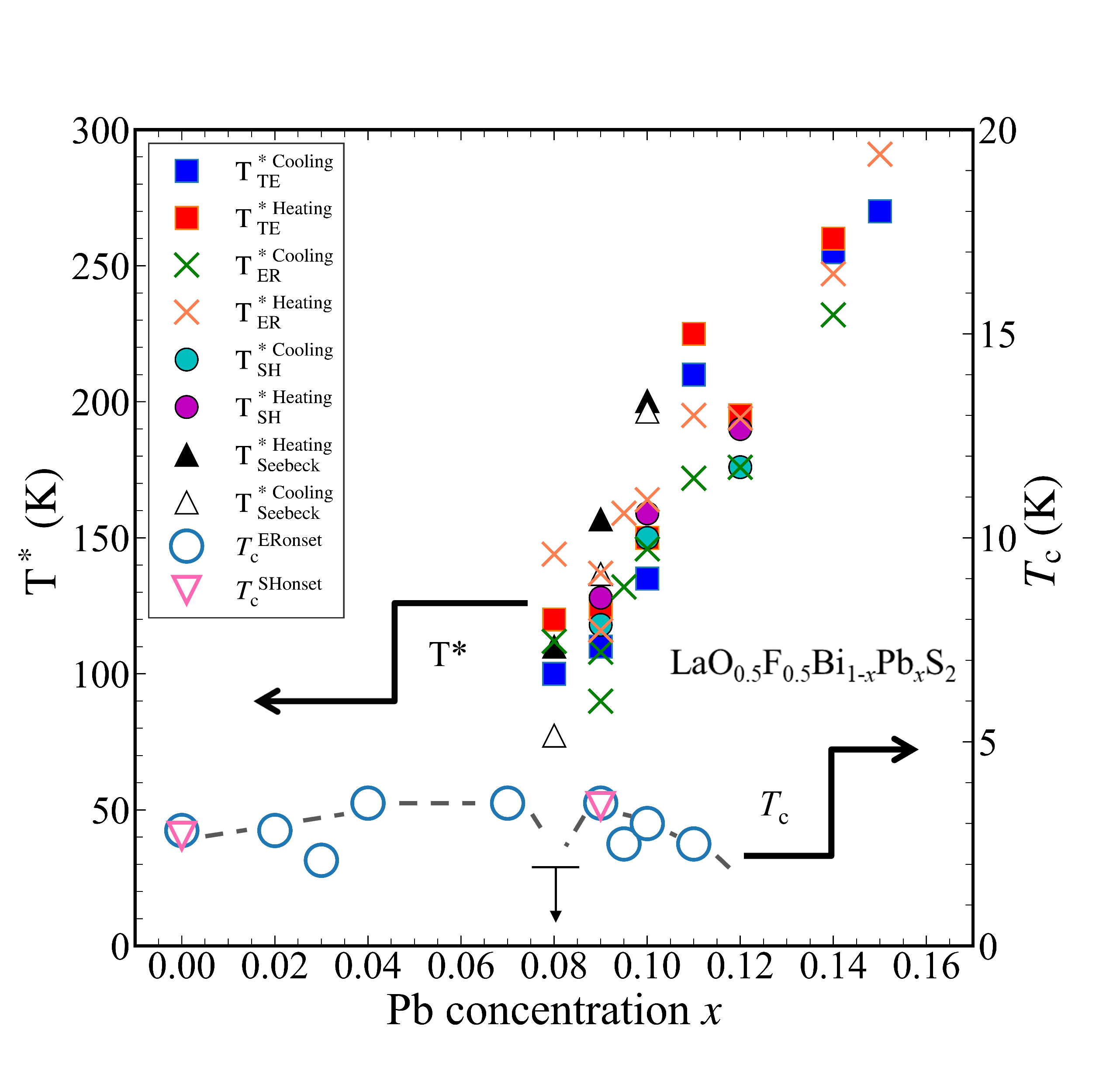}
  \caption{Pb concentration (${x}$) dependence of T* (thermal expansion, electrical resistivity, specific heat, and Seebeck coefficient), and $T_{c}^{ onset}$ (resistivity and specific heat) for  LaO$_{0.5}$F$_{0.5}$Bi$_{1-{x}}$Pb$_{x}$S$_{2}$ (${x}$=0${\sim}$0.15). 
The data for $T_{c}^{ER onset}$ and $T_{c}^{SH onset}$ are shown by cyan circles and pink triangles.
The data for T*$_{TE}$, T*$_{ER}$, T*$_{SH}$ and T*$_{Seebeck}$ are shown by red and blue squares, orange and green crosses, cyan and magenta circles, and black and white triangles in heating and cooling processes, respectively.  }
  \label{f7}
\end{figure}

\end{document}